# Group theory approach to combination tones

Tadeusz Ziębakowski

*Institute of Manufacturing Engineering, Technical University of Szczecin*
*Al. Piastów 19, 70-310 Szczecin, Poland*
*Electronic mail*: harmat@ps.pl

Traditional approaches to combination tones based on Helmholtz theory encounter essential interpreting difficulties, which the most known example is the anomalous behaviour of the combination tone $2\omega_1 - \omega_2$. Without doubt the phenomenon of combination tones is connected with nonlinear phenomena in the human ear. Therefore, it is generally assumed that the explanation of this phenomenon is possible on the level of the ear. This paper proposes different approach, which assumes the existence of an additional mechanism beyond the ear. The aim of this mechanism is to diminish sensitivity of the auditory process to nonlinear distortions by appropriate decoding of sound information. The general rule of construction and the mathematical model of such a mechanism is proposed, which allows for the explanation of many problems associated with combination tones. As a consequence an unexpected result is obtained with regard to a combination tone $2\omega_1 - \omega_2$ that it can originate already from the nonlinear distortion of the second order. This mathematical model is formulated in terms of finite groups.

## I. INTRODUCTION

The problem of nonlinear sound distortion in the auditory process is considered in this work. Many experiments have been carried out that support the existence of considerable nonlinear distortions of the sound inside the human ear. Moreover, the type of these distortions differs significantly among people. Meanwhile, such a strong influence of nonlinear distortion on auditory sensations is not observed (Boer, 1991, Sects. 5.3 and 8.3) and thereby auditory sensations to considerably less degree depend on a person. Therefore, we postulate existence of such an additional mechanism in the process of the auditory perception that diminishes our sensitivity to nonlinear distortions. This mechanism may rely for example on a suitable decoding of the sound. According to this hypothesis, we will construct some kinds of quasi-invariants with respect to nonlinear mechanical transformations where the latter is the source of distortion. Let us first consider a notion of the nonlinear mechanical transformation (NMT).

From the point of view of information transfer, it is important to separate the intermediate stages associated with the medium of sound propagation. The transition from one stage to another we will call transformation. Since these transformations are of a mechanical nature, they can be described using the principles of dynamics. Hence the class of such transformations is described in general by nonlinear differential operators. We consider the subclass of such operators, which are obtained by polynomial functions over linear differential operators. Although these operators do not allow to describe accurately many real transformations they can still form a good approximation. These operators will be identified with NMT's [1]. The degree of polynomial determines the order of nonlinearity. From the spectral analysis point of view NMT's are sources of significant distortions, namely the spectrum of a sound acquires many new components called *nonlinear tones*. For example, if we have a signal composed by two tones of frequencies $\omega_1, \omega_2$ then NMT's of second order introduce nonlinear tones with the following frequencies: $2\omega_1$, $2\omega_2$, $\omega_1 + \omega_2$, $\omega_1 - \omega_2$ and in the third order: $3\omega_1$, $3\omega_2$, $2\omega_1 + \omega_2$, $2\omega_1 - \omega_2$, $2\omega_2 - \omega_1$, $2\omega_2 + \omega_1$. Such nonlinear phenomena cause many difficulties in understanding the perception of pitch. In 19th century Ohm formulated the law that the particular components of the spectrum are perceived via their pitch. Almost simultaneously, there appeared the problem of combination tones that are heard as additional pitches (audible tones) in signals consisting of two frequency components. Helmholtz (1856,1863) interpreted these combination tones as nonlinear tones since they have compatible frequencies. However, such an interpretation of hearing of combination tones leads to many peculiarities listed below:

1. The number of combination tones is considerably lower than the number of possible nonlinear tones. Moreover, they appear at the determined frequency ratios of primary tones (Plomp, 1965; Smoorenburg, 1972a).
2. The most prominent combination tones are the quadratic difference tone $\omega_2 - \omega_1$ and the higher-order combination tones of the type $\omega_1 - n(\omega_2 - \omega_1)$, with small integer number $n$ and $\omega_2 > \omega_1$. These latter are audible at a relatively low level of the primary tones. It is interesting that the 3$^{rd}$-order tone $2\omega_1 - \omega_2$ ($n = 1$) is audible at lower levels of the primary tones in comparison to the 2$^{nd}$-order tone $\omega_2 - \omega_1$ (Plomp, 1965, 1976; Smoorenburg, 1972a).

---

[1] Let $T$ be a real polynomial of $n+1$ variables. We will describe NMT as a mapping of signal $x(t)$ into signal $y(t)$ by operator $\widetilde{T}: x \mapsto y = T(x, \frac{dx}{dt}, \frac{d^2x}{dt^2}, \dots \frac{d^nx}{dt^n})$. In many real cases a starting signal $x(t)$ and signal $y(t)$ obtained as a result of trans-

formation are related by equation $x(t) = \widetilde{S}(y(t))$ where $\widetilde{S}$ is polynomial differential operator. Consequently, such a transformation is described by operator $\widetilde{T} = \widetilde{S}^{-1}$, which is not in general a polynomial differential operator (see Helmholz, 1863, App. XII).

3. The tones $2\omega_1 + \omega_2$ and $2\omega_2 + \omega_1$ are not audible and thereby do not accompany the audible tone $2\omega_1 - \omega_2$. Moreover, the tone $2\omega_2 - \omega_1$ is not audible at lower signal level. (Plomp 1965).

4. If amplitudes of $\omega_1$ and $\omega_2$ are both equal to some value $x$ then $2\omega_1 - \omega_2$ as a nonlinear tone should increase with $x^3$ at low $x$. On the contrary, experiments show that it increases almost with the first power of $x$ (Zwicker, 1955, 1968; Goldstein, 1967; Helle, 1969; Smoorenburg, 1972b; Zwicker, Fastl, 1973).

These problems were also specified by Boer (1984, Sect. 1). It is the purpose of this work to explain the aforementioned problems in a unified way. The Ohm law associates correctly the pitch sensation with the sound spectrum but singularities in hearing of the combination tones may lead to the conclusion that these two categories should not be directly identified. This paper assumes the existence of the additional mechanisms in the high level of the auditory process that follows the spectral analysis in the human ear. The aim of such mechanisms is to select (filter) such sound information, which is low sensitive to distortions caused by NMT's. As a result we obtain a new image of the sound that is less susceptible to nonlinear distortions and also reflects the world of our auditory sensations. Therefore, this level of the auditory process we will call sensation analysis. We will assume that this process is determined by its symmetry. In our considerations we take into account the fact that in auditory process a continuous sound signal is estimated by a finite set of parameters. Therefore, rather than search for quasi-invariants with respect to NMT's we will search for groups, which in some way simulate NMT's on the set of these parameters and search for invariants under such groups. We will postulate the relationship of these invariants with the components of the auditory sensations such as pitch, consonances, timbres and phonemes. Since the phenomenon of combination tones concerns the pitch perception, we concentrate only on the pitch in further part of this paper. We will show that thanks to this construction it is possible explains the aforementioned peculiarities.

## II. OUTLINE OF THEORY

Sensation analysis, like spectral analysis, may be described as a transformation, which maps one characteristics into other one. As a initial characteristics we assume a certain discrete spectral characteristics $F$ of sound signal e.g. it's discrete Fourier transform. Our aim is to construct some new characteristics called *sensation characteristics*, which is a result of this transformation. At first, we show general construction of such a characteristics and next, we outline structure of space of elementary sensations, which is a domain of such a characteristics.

### A. Sensation characteristics

To begin, we will introduce some construction elements. By $K$ we denote a discrete set of spectral frequencies, both the positive and the negative, thus spectral characteristic is a mapping $F : K \to C$. We assume for simplicity that spectrum components are "normalized" i.e. the auditory threshold value is $|F(\omega)| = 1$, for each $\omega \in K$. Moreover, we assume that for the sine signal in the form $A\sin(\omega t + \varphi)$ is satisfied $|F(\omega)| = |F(-\omega)| = A$ (without factor ½). We will also use modified logarithmic function:

$$\log^+(x) := \begin{cases} \log(x) & \text{for} \quad x \geq 1 \\ 0 & \text{for} \quad 0 \leq x < 1 \end{cases} \qquad (1)$$

Let us consider a "n-dimensional" frequency domain $W = K^n$ i.e. n-th cartesian power of the set $K$. Furthermore, let us consider a set $S$ of some disjoint subsets of the space $W$. The elements of the set $S$ we will interpret as elementary auditory sensations such as pitch, consonances, timbres or phonemes, so set $S$ will be also called *sensation domain*. Having spectral characteristics $F$ we define a certain real measure $F^\#$ on $W$, i.e. for $S \subset W$:

$$F^\#(S) := \sum_{[\omega_1, \omega_2, \ldots, \omega_n] \in S} \log^+ \left\| F(\omega_1) F(\omega_2) \cdot \ldots \cdot F(\omega_n) \right\| \qquad (2)$$

where $\|\cdot\|$ is some "norm" of complex number for example the absolute value which is denoted by $|\cdot|$. We will use this meaning of the symbol $\|\cdot\|$ in our considerations, although we will also suggest another possibility in Sect. VI.

In practice we are interested in the restriction of the measure $F^\#$ to $S$, which we will call a *sensation characteristics* and we will also denote it by $F^\#$. We will associate a psychoacoustic interpretation to $F^\#$. Namely, the characteristics $F^\#$ allows us to answer the question "how much" of each elementary sensations described by set $S$ one can hear in the sound with the spectrum $F$. Let us note that the definition of $F^{\#}$ takes into account Weber-Fechner law due to the use of modified logarithmic function. Thus $F^{\#}(S)$ expresses the level of elementary sensation $S \in S$. The above construction is shown schematically in Fig. 1 and is explained by the following example, which concerns the problem of pitch of complex tone.

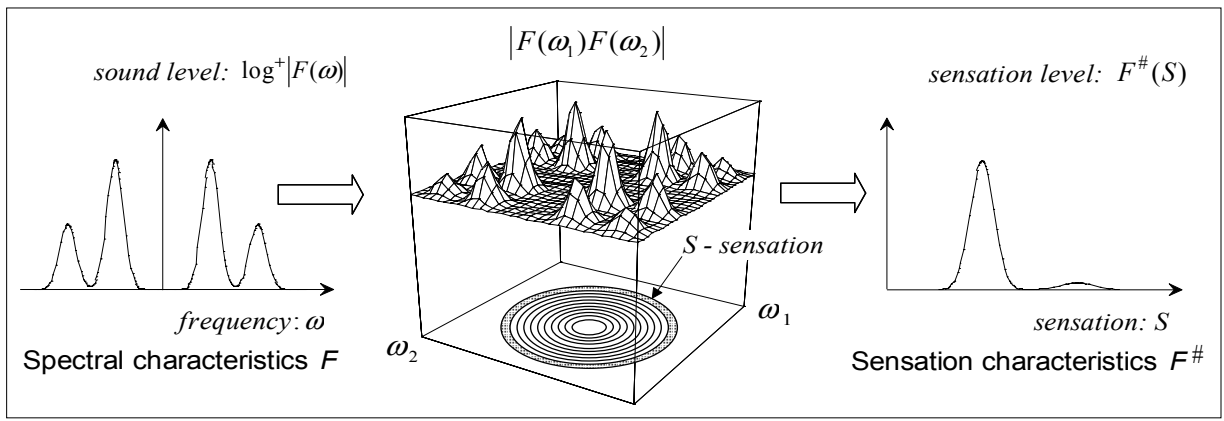

Fig. 1. Scheme of construction of the sensation characteristics.

## B. Example 1

Let us consider the case of $K = Z$ where $Z$ is the ring of integer numbers[2], and $W = Z \times Z$. Let $Z^{+}$ be the set of positive integer numbers. As the sensation domain $S$ let us assume a family of subsets $\{P_{\omega}\}_{\omega \in Z^{+}}$ defined in the following way:

$$P_{\omega} := \{[\omega_1, \omega_2] \in W : GCD(\omega_1, \omega_2) = \pm\omega\} \quad (3)$$

where $GCD(\omega_1, \omega_2)$ denotes the greatest common divisor of numbers $\omega_1, \omega_2$. The subset $P_{\omega}$ is interpreted as pitch of the tone with frequency $\omega$. For clarity let us write out some elements of subset $P_{\omega}$:

$$
\begin{aligned}
P_{\omega} = \{ & [\pm\omega, \pm\omega], [\pm\omega, \pm 2\omega], [\pm\omega, \pm 3\omega], [\pm\omega, \pm 5\omega], \\
& [\pm 2\omega, \pm 3\omega], [\pm 2\omega, \pm 5\omega], [\pm 3\omega, \pm 5\omega], [\pm 2\omega, \pm\omega], \\
& [\pm 3\omega, \pm\omega], [\pm 5\omega, \pm\omega], [\pm 3\omega, \pm 2\omega], [\pm 5\omega, \pm 2\omega], \\
& [\pm 5\omega, \pm 3\omega], \dots\}
\end{aligned}
\quad (4)
$$

where $\omega = 1, 2, 3, \dots$. Let us now consider an example spectral characteristics $F$ describing a complex tone. We write it in the following form:

$$F = (\dots, F_{-l}, F_{-l+1}, \dots, F_{-1}, F_0, F_1, \dots, F_{l-1}, F_l, \dots) \quad (5)$$

where $F_i \in C$ and $F_{-i} := \overline{F_i}$ and for some chosen $\Omega \in Z^{+}$ as a frequency of the fundamental we set:

| | |
|---|---|
| $\|F_{\pm\Omega}\| = 10^5$ (50dB) | $\|F_{\pm 2\Omega}\| = 10^4$ (40 dB) |
| $\|F_{\pm 3\Omega}\| = 10^3$ (30 dB) | $\|F_{\pm 4\Omega}\| = 10^2$ (20 dB) |
| $\|F_{\pm 5\Omega}\| = 10$ (10 dB) | remaining $F_i$ are equal to 0 |

Furthermore, let us compute $F^{\#}$ on the sets $P_{\omega}$ according to Eq. (2). Upper chart in Fig. 2, plotted for the above data, shows that characteristics $F^{\#}$ have the property of amplifying the fundamental tone and drowning harmonics. After the elimination of the fundamental tone, the so-called residual (low) pitch remains, which originates from harmonics in the complex tone. This situation is shown in the lower chart in Fig. 2, which is obtained under the assumption $\|F_{\pm\Omega}\| = 0$. In practice this effect is not as strong as it is shown in Fig. 2b. It appears at the signal level of about 50 dB. (Plomp 1965, 1976 Chap. 7).

## C. Sensation domain

This simple example shows that by proper choice of the sensation domain $S$, i.e. subsets on $W$ representing elementary auditory sensations, the characteristics $F^{\#}$ diminishes the influence of these components, which correspond to nonlinear components of the characteristics $F$. We now consider this in more details.

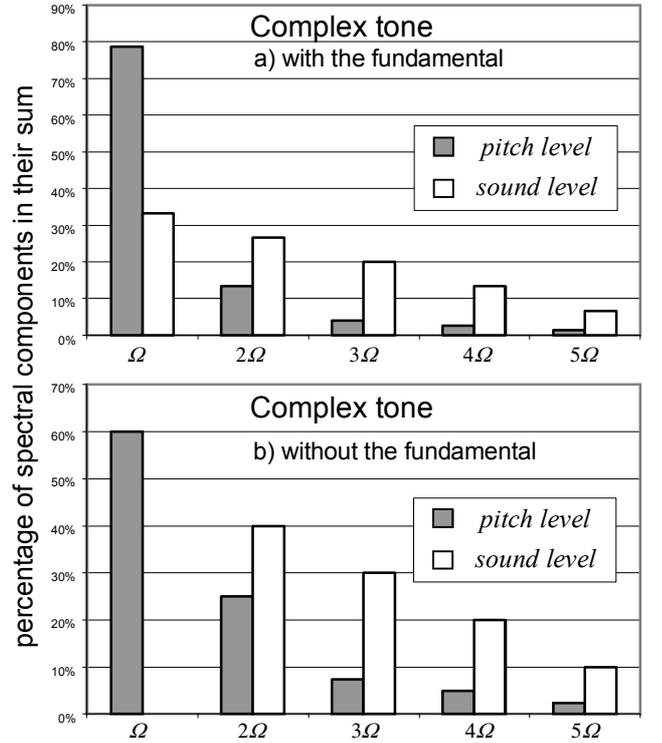

Fig. 2. The residual pitch effect. The sound level is given by $\log^+(|F(\omega)|)$ and the pitch level is given by $F^{\#}(P_\omega)$. In practice residual pitch effect is not as strong as it is shown in b).

The ring of integer numbers is not convenient to enumerate spectral components. First, it is the infinite set while the sound information received by the human brain is finite; second, there are many more problems in consideration of matrix groups over rings than over fields. Therefore, to enumerate components of the discrete sound spectrum we use the appropriately large simple finite field $K = Z_p$ i.e. the field of integers modulo $p$, where $p$ is a prime number[3]. Let us introduce the ordering of this field according to the sequence:

$$-l, -l+1, \dots, -1, 0, 1, \dots, l-1, l \quad (6)$$

where $l$ is such that $2l + 1 = p$. If $K$ is a field then set $W$ has structure of a $n$-dimensional linear space over field $K$. In this case the definition of $F^{\#}$ given by Eq. (2) is connected with choice of a certain basis, which will be called spectral basis. In such basis, coordinates have meaning of the frequencies of spectral components. In order to set the sensation domain $S$ i.e. to distinguish subsets on $W$, we introduce some matrix group $H$ on $W$. The group $H$ divides space $W$ into subsets called orbits, and we assume that these orbits form domain $S$. The set of all orbits is usually denoted by $W/H$ so that $S = W/H$. With regards to the group $H$, we will assume that such a group will in some way simulate NMT's. Let us try to answer the question: what does it mean that a group simulates NMT's.

Let $u = \{u_i\}_{i=1\dots n}$ and $v = \{v_i\}_{i=1\dots n}$ are vectors of the space $W$ with coordinates in a spectral basis. We say that $v$ is deriva-

---

[2] Since the sum in the definition of $F^{\#}$ given by Eq. (2) must be finite, so we assume that $F$ has a finite support in the case $K = Z$.

[3] It is usually assumed that there are about 1400 components, so $p$ should be greater than this number.

*tive vector of order k with respect to u* if every coordinate $v_i$ can be obtained as a linear combination of coordinates $u_j$ with "small integer numbers" i.e. $v_i = \alpha_1 u_1 + \alpha_2 u_2 + \ldots + \alpha_n u_n$ where $\alpha_i = -k, -k+1, \ldots, k-1, k$ and $|\alpha_1| + |\alpha_2| + \ldots + |\alpha_n| \leq k$. Thus, simulation of NMT's by the group $H$ relies on generation and collection of derivative vectors into orbits. We will illustrate this by the next example.

### D. Example 2

Let $W$ be 2-dimensional linear space over field $K = Z_7$ and let us choose a spectral basis $\{e_1, e_2\}$. Let $H$ be the group generated by the matrices:

$$\lambda := \begin{bmatrix} 1 & -1 \\ 1 & 0 \end{bmatrix} \quad \sigma := \begin{bmatrix} 0 & 1 \\ 1 & 0 \end{bmatrix} \tag{7}$$

It can be checked that $\lambda^6 = 1$, $\sigma^2 = 1$, $\lambda\sigma = \sigma\lambda^5$ so that the group $H$ consists of 12 elements[4]. According to definition of the spectral basis, if we treat coordinates in basis $\{e_1, e_2\}$ as frequency values, then group $H$ simulates NMT's of the second order by generating some 2nd order derivative vectors, e.g.:

$$\lambda \begin{bmatrix} \omega_1 \\ \omega_2 \end{bmatrix} = \begin{bmatrix} \omega_1 - \omega_2 \\ \omega_1 \end{bmatrix} \qquad \lambda \begin{bmatrix} \omega_1 \\ -\omega_2 \end{bmatrix} = \begin{bmatrix} \omega_1 + \omega_2 \\ \omega_1 \end{bmatrix}$$

$$\lambda \begin{bmatrix} \omega_1 \\ -\omega_1 \end{bmatrix} = \begin{bmatrix} 2\omega_1 \\ \omega_1 \end{bmatrix} \tag{8}$$

Group $H$ establishes the set of following 8 orbits on $W$ which form the space $W/H$:

$$O := \{[0, 0]\}$$
$$A_\omega := \{[-\omega, -\omega], [-\omega, 0], [0, -\omega], [0, \omega], [\omega, 0], [\omega, \omega]\}$$
$$B_\omega := \{[-2\omega, -\omega], [-\omega, -2\omega], [-\omega, \omega], [\omega, -\omega], [\omega, 2\omega], [2\omega, \omega]\}$$
$$C := \{[-3, -2], [-3, -1], [-2, -3], [-2, 1], [-1, -3], [-1, 2],$$
$$[1, -2], [1, 3], [2, -1], [2, 3], [3, 1], [3, 2]\} \tag{9}$$

where $\omega = 1, 2, 3$. If we interpret parameter $\omega$ as frequency, we may say that orbits $A_\omega$ and $B_\omega$ are covariant with respect to frequency, namely:

$$k A_\omega = A_{k\omega}, \qquad k B_\omega = B_{k\omega} \tag{10}$$

The orbit $C$ is invariant with respect to frequency because $kC = C$ for $k \in K \setminus \{0\}$. The orbit $O$ is trivial and will not be considered any further. We use nontrivial orbits to characterise elementary auditory sensations. Shown below is an example of an abstract psychoacoustic interpretation:

$$A_\omega \leftrightarrow \text{pitch}$$
$$B_\omega \leftrightarrow \text{octave consonance}$$
$$C \leftrightarrow \text{any phoneme or timbre}$$

It turns out however, that if the group $H$ generates "too many" derivative vectors then the sensation domain contains "too few" orbits to map auditory sensations. For example, if group generates all derivative vectors of the 2nd order with respect to each vector then space $W$ is trivially divided into two orbits: $\{\mathbf{0}\}$ and $W \setminus \{\mathbf{0}\}$. Generally, there is large arbitrariness to create groups, which simulate NMT's in the given meaning. Therefore we have to search for additional arguments to determine group $H$ and at the same time structure of the sensation domain.

## III. BASIC ASSUMPTIONS AND PSYCHOACOUSTICAL POSTULATES

In this section we determine more precisely construction outlined in previous section. The main role of group $H$ in the presented theory is to simulate NMT's by collecting derivative vectors into orbits. However, as mentioned, this feature of group $H$ does not allow to determine it's completely. Moreover, we have to establish another parameters of our construction i.e. the field $K$ and dimension of the space $W$. To this end we use musical and physiological arguments. Since the sensation domain $W/H$ reflects directly the image of our auditory sensations we will try to say something about structure of this domain on the basis of our hearing experience. It will be formulated in the form of psychoacoustical postulates, which will allow us to determine the group $H$.

### A. Field K and dimension of space W

At first, let us consider the problem of compatibility between the ordering structure and the algebraic structure of the field. Pitch and frequency are categories, which have natural ordering. However, in the finite field one cannot introduce a linear ordering structure compatible to an algebraic structure[5]. The ordering defined earlier in Section II has some disadvantages e.g. the sum and product of positive elements could be negative. To avoid violation of the ordering structure by algebraic operations we will narrow the acceptable range of numbers.

Using the music theory arguments we will determine a field $K = Z_p$ i.e. we will establish prime number $p$. Frequency doubling is connected with special similarity of music impression. For example, it is expressed in the structure of music scales and distinctive role of octave consonance. We can separate about nine octaves from the auditory area, therefore we should expect that

$$1 < 2 < 2^2 < \ldots < 2^9 \tag{11}$$

Since the field is finite, there exists a minimal integer number $q$ such that $2^q \equiv 1 \pmod{p}$, which means that for some integer $s$ equation $2^q = sp + 1$ is satisfied. To avoid situation when there is another power of number 2 between two consecutive powers of number 2, it is sufficient to assume $s = 1$ hence $p = 2^q - 1$. For $Z_p$ to be a field, the number $p$ must be a prime. Only for some prime numbers $q$ e.g. 13,17,19,31 etc. correct values of $p > 2^9$ are obtained. Although most of the presented results do not depend on the choice of a finite field we will use the field $Z_p$ where $p = 2^{13} - 1 = 8191$ in the further calculations.

---

Now we consider the problem of choosing of the dimension of the space $W$. For this purpose we will use physiological arguments. Sound information arriving to the brain comes from four or more rows of hair cells that are located in the cochlea. We assume the hypothesis that each row of hair cell in the cochlea provides spectrum information to the particular coordinate in the spectral basis of the space $W$. There are four rows along the full length of the cochlea therefore we assume that $W$ is a four-dimensional linear space over $K$ which is isomorphic to $K^4$. We will return to this topic in the subsection IV.E.

### B. Multiplication of orbits by scalars

If we multiply all elements of orbit $O \in W / H$ by $k \in K$ (we will write $k \cdot O$) then as a result we obtain either the same orbit or another orbit. If we consider set of scalars $K_O$ preserving given orbit $O$, then it forms a group which is the subgroup of the group containing all invertible elements of the field $K$ denoted by $K^*$. We say that $K_O$ is the *scalar group of orbit $O$*. For example, in the field $K = Z_{8191}$, if numbers $-1$ and $2$ preserve the given orbit $O$, then the scalar group contains the following 26 elements: $\{\pm1, \pm2, \pm4, \pm8, \pm16, \pm32, \pm64, \pm128, \pm256, \pm512, \pm1024, \pm2048, \pm4096\}$. We will write $< g_1, g_2, ..., g_k >$ to denote group generated by $g_1, g_2, ..., g_k$, so $K_O = <-1, 2>$ and let us note that $K_O = <-2>$. Let us consider the orbit $O$ with nontrivial scalar group $K_O$ and let $X$ denote a set of representatives of cosets $K^*/K_O$. Then such an orbit determines a family of orbits $\{O_\omega\}_{\omega \in X}$ with scalar group $K_O$ in the following way: $O_\omega := \omega \cdot O$ for $\omega \in X$. We say that the family $\{O_\omega\}_{\omega \in X}$ is generated by orbit $O$ with scalar group $K_O$. The above construction can be described in a different manner. The group $K^*$ acts in a natural way on the space $W$, on which the group $H$ also acts. Since $K^*$ commutes with $H$, $K^*$ acts also on $W/H$. Therefore, the family of orbits $\{O_\omega\}_{\omega \in X}$ is $K^*$ orbit on $W/H$ and the scalar group $K_O$ is simply the stability group of element $O$. Since $K^*$ is abelian, the stability group of each orbit $O_\omega$ is isomorphic to $K_O$. Therefore group $K_O$ is a property of the family $\{O_\omega\}_{\omega \in X}$.

### C. Psychoacoustical postulates determining group H

Before we proceed to formulation of psychoacoustical postulates, we will introduce a few definitions. Let $X$ and $Y$ are sets of representatives of cosets $K^*/<-1>$ and $K^*/<-2>$ respectively. It is convenient to assume that $X$ is the set $K^+$ of all positive elements of the field $K$. Having chosen spectral basis, we define $E$ as the set of vectors whose coordinates have values $0, 1, -1$. This set contains vectors "with pattern frequency" which is equal to 1. Similarly, we define $D$ as the set of vectors whose coordinates have values from coset $< -2 >$ and $0$ (i.e. $0, \pm1, \pm2, \pm4$ etc.). We now present basic assumptions of our theory of sensation analysis in the form of the following postulates.

*Postulates concerning the pitch:*

P1.    The sensation domain $W/H$ contains exactly one family of orbits $\{P_\omega\}_{\omega \in X}$ which is generated by an orbit with the scalar group $< -1 >$. Each orbit will be interpreted as pitch corresponding to frequency $\omega$.

P2.    Among the orbits of the family $\{P_\omega\}_{\omega \in X}$ exactly one orbit nontrivially intersects the set $E$. By proper adjustment of parameter $\omega$ we will assume that it is the orbit $P_1$. In this way the mapping frequency - pitch: $X \ni \omega \mapsto P_\omega = \omega P_1 \in W / H$ is well defined.

*Postulates concerning the octave consonance:*

O1.    The sensation domain $W/H$ contains exactly one family of orbits $\{O_\xi\}_{\xi \in Y}$ which is generated by an orbit with the scalar group $< -2 >$. Such orbits will be interpreted as octave consonances.

O2.    Among the orbits of the family $\{O_\xi\}_{\xi \in Y}$ exactly one orbit nontrivially intersects the set $D$. In this way the connection between octave consonance and frequency is well established.

*Postulate concerning remaining components of auditory sensation:*

T.    The remaining orbits of the sensation domain $W/H$ are $K^*$ invariant. We have in mind here such elementary sensation as timbre, non-octave consonance and phoneme.

In another words, the latter sensations are invariant with respect to frequency. Postulate P1 means that pitch is covariant with respect to frequency. Postulate O1 means that octave consonance is invariant with respect to frequency doubling. Postulates P2 and O2 calibrate pitch and octave consonance with respect to frequency. Let us also note that postulates P1 and P2 reformulate the Ohm's acoustical law. Above postulates are shown schematically in Fig. 3.

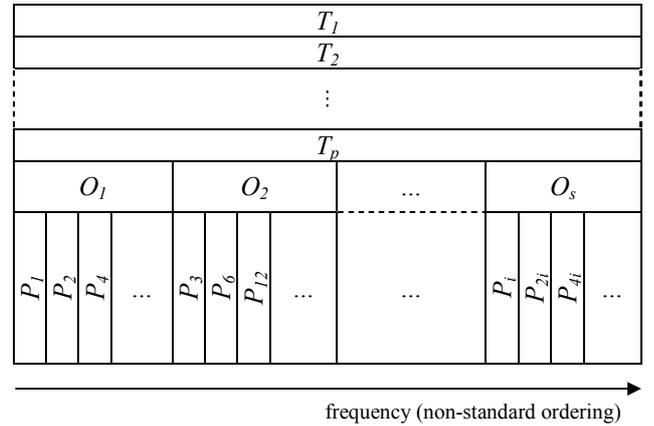

Fig. 3. Partition scheme of space $W\backslash\{0\}$ by $H$-orbits and psychoacoustic interpretation of the sensation domain $W/H$. Pitch orbits are denoted by $P_1, P_2, P_3, ...$, octave consonance orbits by $O_1, O_2, O_3, ...$. Orbits representing sensation independent on frequency e.g. timbre, non-octave consonance, phoneme, are denoted by $T_1, T_2, T_3, ...$.

Taking into account considerations from previous section with regards to sensation characteristics we should complete above postulates by the following:

*Postulate concerning the measurement of sensations:*

M.    The measurement of auditory sensations is determined by a mapping, denoted by symbol $^\#$, which assigns a finite real measure $F^\#$ on $W$ to any spectral characteristics $F$. The level of elementary sensation $S \in W / H$ in the sound with spectrum $F$ is determined by $F^{\#}(S)$. The restriction of the measure $F^{\#}$ to $W/H$ is called sensation characteristics.

The example of such a mapping is given in Eq. (2). This Equation can also be written in a different form by using tensor product operation:

$$F^{\#}(S) = \sum_{w \in S} \log^+ \left\| F \otimes F \otimes \ldots \otimes F(w) \right\| \qquad (12)$$

In Section IV.E we will also create other "one-dimensional" spectral characteristics $F_1, F_2, \ldots, F_n$ on the base of characteristics $F$ e.g. by restriction to frequency range and we will associate them with elements of the spectral basis. Therefore, we will use generalization of the Eq. (12) in the form:

$$F^{\#}(S) := \sum_{w \in S} \log^+ \left\| F_1 \otimes F_2 \otimes \ldots \otimes F_n(w) \right\| \qquad (13)$$

It should be emphasized that both definitions (12) and (13) require introduction of the spectral basis.

## IV.    EXAMPLE MODEL

We will present now the model of sensation analysis, which satisfies postulates of our theory. The additional simplifications will be helpful in understanding our method. However, predictions of this model will only be partially consistent with experiment. After introducing group $H$ and division of the space $W$ into orbits, we will focus our attention to determining pitch orbits. For this purpose we will construct a useful tool, namely $H$-quasi-invariant quadratic form. Furthermore, we will adjust # mapping. This model will help us to present a solution of the problem of combination tones, which is considered in the next section.

### A.    Group H

In our construction we have already selected the spectral basis in space $W$ in which is defined $F^{\#}$. Let us denote this basis by $B = \{e_i\}_{i=1..4}$. We introduce an additional auxiliary basis $C = \{f_i\}_{i=1..4}$ to define the group and later we will return to basis $B$ by appropriate transfer matrix. We will work on the field $\mathbf{K} = \mathbf{Z}_p$ where $p = 8191$ and we will choose an element $\lambda \in \mathbf{K}$, which generates the cyclic group $\mathbf{K}^*$. Let us consider the following matrices in the basis $C$

$$k := \begin{bmatrix} 1 & & & \\ & -1 & \mathbf{0} & \\ & \mathbf{0} & 2 & \\ & & & \frac{1}{2} \end{bmatrix}, l := \begin{bmatrix} \frac{1}{\lambda} & & & \\ & \lambda & \mathbf{0} & \\ & \mathbf{0} & \lambda & \\ & & & \lambda \end{bmatrix},$$

$$m := \begin{bmatrix} 1 & 0 & 0 & 1 \\ & 1 & 0 & 0 \\ & & 1 & 0 \\ \mathbf{0} & & & 1 \end{bmatrix}, n := \begin{bmatrix} 1 & 0 & 0 & 0 \\ 0 & 1 & 0 & 2 \\ 0 & 2 & 1 & 2 \\ 0 & 0 & 0 & 1 \end{bmatrix} \qquad (14)$$

We will be interested in the group $H$, which is generated by above four matrices. Elements $n$ and $m$ are commutative and each of them generates cyclic subgroup of the rank 8191 in the following forms:

$$M := \left\{ \begin{bmatrix} 1 & 0 & 0 & \mu \\ & 1 & 0 & 0 \\ & \mathbf{0} & 1 & 0 \\ & & & 1 \end{bmatrix} : \mu \in K \right\}$$

$$N := \left\{ \begin{bmatrix} 1 & 0 & 0 & 0 \\ 0 & 1 & 0 & 2\nu \\ 0 & 2\nu & 1 & 2\nu^2 \\ 0 & 0 & 0 & 1 \end{bmatrix} : \nu \in K \right\} \qquad (15)$$

Matrix $k$ generates the subgroup $K$ of the rank 26, which is isomorphic to the group $Z_2 \times Z_{13}$ while matrix $l$ generates the subgroup $L$ of the rank 8190, which is isomorphic to $\mathbf{K}^*$. From the form of matrices $k, l, m, n$ one concludes that the subgroups $KL$ and $MN$ are abelian. It can also be checked that the subgroup $MN$ is a normal subgroup of $H$ and $H$ is a semidirect product of the groups $MN$ and $KL$, so that $H = KLMN$. We will see in a moment that such construction allows one to obtain the structure of orbit, which was postulated in the previous section.

### B.    Division into orbits

Now we describe the sensation domain $W/H$ by giving representative of the particular orbit. According to postulate P1 space $W/H$ contains the following unique family of orbits, which is generated by an orbit with the scalar group $< -1 >$.

$$P_\omega := H \begin{bmatrix} 1 \\ \omega^2 \\ 0 \\ 0 \end{bmatrix} \qquad \text{for} \quad \omega \in \mathbf{K}^+ \qquad (16)$$

Let $Y$ denote as in III.C the set of representatives of cosets $K^* / < -2 >$. It can be checked that the space $W/H$ contains also another unique family of orbits which is generated by an orbit with the scalar group $< -2 >$:

$$O_\xi := H \begin{bmatrix} 1 \\ 0 \\ \xi \\ 0 \end{bmatrix} \qquad \text{for} \quad \xi \in Y \qquad (17)$$

All remaining orbits are invariant with respect to frequency and are generated by the following vectors:

$$\begin{bmatrix} 0 \\ 0 \\ 0 \\ 0 \end{bmatrix}, \begin{bmatrix} 1 \\ 0 \\ 0 \\ 0 \end{bmatrix}, \begin{bmatrix} 0 \\ 1 \\ 0 \\ 0 \end{bmatrix}, \begin{bmatrix} 0 \\ 0 \\ 1 \\ 0 \end{bmatrix}, \begin{bmatrix} 0 \\ 0 \\ 0 \\ 1 \end{bmatrix}, \begin{bmatrix} 0 \\ 0 \\ 1 \\ \xi \end{bmatrix} \qquad \text{for} \quad \xi \in Y \quad (18)$$

Thus postulates O1 and T are also satisfied.

## C.  H-invariant subspace $W_0$

By $W_0$ we denote the subspace of $W$ spanned by vectors $f_1, f_2, f_3$, i.e.:

$$W_0 := \{xf_1 + yf_2 + zf_3 : x, y, z \in K\} \qquad (19)$$

Let us see, how generators $k, l, m, n$ of the group $H$ act on the space $W_0$:

$$k \begin{bmatrix} x \\ y \\ z \\ 0 \end{bmatrix} = \begin{bmatrix} x \\ -y \\ 2z \\ 0 \end{bmatrix}, \quad l \begin{bmatrix} x \\ y \\ z \\ 0 \end{bmatrix} = \begin{bmatrix} x/\lambda \\ \lambda y \\ \lambda z \\ 0 \end{bmatrix},$$

$$m \begin{bmatrix} x \\ y \\ z \\ 0 \end{bmatrix} = \begin{bmatrix} x \\ y \\ z \\ 0 \end{bmatrix}, \quad n \begin{bmatrix} x \\ y \\ z \\ 0 \end{bmatrix} = \begin{bmatrix} x \\ y \\ z+2y \\ 0 \end{bmatrix} \qquad (20)$$

We observe that $W_0$ is $H$-invariant subspace of the space $W$. Let us also note, that by virtue of Eqs. (16) and (17) orbits $\{P_\omega\}$ and $\{O_\xi\}$ are included in $W_0$.

## D.  H-quasi-invariant quadratic form on $W_0$

Let $q$ be a quadratic form on $W$ given by

$$q([x, y, z, s]) := xy \qquad (21)$$

Equation (20) implies that group $H$ preserves, up to the sign, form $q$ on $W_0$ i.e. form $q$ is invariant up to the sign on orbits included in $W_0$. Moreover Eq. (16) implies that

$$q(P_\omega) = \{\pm \omega^2\} \qquad (22)$$

thus $q$ determines one-to-one mapping between family of pitch orbits and the cosets $K^*/<-1>$. With the help of quadratic form $q$, we can easily check whether a given vector of subspace $W_0$ belongs to a pitch orbit and if so, to which one[6].

## E.  # mapping

According to the assumptions of subsection III.A each row of the hair cells provides spectral characteristics. These spectral characteristics are one-dimensional. In the definition of $F^\#$ in Eq. (13) each characteristics is associated with certain coordinates in the spectral basis denoted by $B$. Thus coordinates in this basis will be interpreted as spectral frequencies. In subsection A we also introduced the auxiliary basis $C$ in which, we defined the group $H$. Let us denote coordinates in basis $B$ by $[u, v, w, t]$ and in basis $C$ by $[x, y, z, s]$. We will give a priori a transfer matrix $T : [u, v, w, t] \to [x, y, z, s]$, which modifies matrix generators of group $H$ in such way that allows one to generate and collect derivative vectors by the group $H$ sufficiently well.

---

[6] The field $\boldsymbol{K}$ does not contain primitive $4^{\text{th}}$ roots of unity. Thus, if $k \in \boldsymbol{K}^*$ then either equation $\omega^2 = k$ or equation $\omega^2 = -k$ has solutions with respect to $\omega$.

$$T := \begin{bmatrix} 1 & 1 & 0 & 1 \\ -1 & -1 & 0 & 1 \\ 1 & 1 & 1 & -1 \\ 2 & 1 & 1 & -1 \end{bmatrix} \qquad (23)$$

The choice of such form of matrix is in agreement with postulates P2 and O2 (we will check postulate P2 in a moment) and has been also successfully tested on the pitch of complex tone problem (see subsection II.B). Let us note that in the basis $B$, subspace $W_0$ is specified by the condition:

$$2u + v + w - t = 0 \qquad (24)$$

whereas the quadratic form $q$ from Eq. (21) has form:

$$q([u, v, w, t]) = t^2 - (u+v)^2 \qquad (25)$$

In order to check postulate P2, it is sufficient to check the value of the form $q$ on the set $E \cap W_0 \supset E \cap \bigcup_{\omega \in X} P_\omega$. Its elements are shown below, together with the corresponding values of the form $q$:

$$\pm \begin{bmatrix} 0 \\ -1 \\ 1 \\ 1 \end{bmatrix}, \quad \pm \begin{bmatrix} 0 \\ 0 \\ 0 \\ 1 \end{bmatrix}, \quad \pm \begin{bmatrix} 1 \\ -1 \\ 0 \\ 1 \end{bmatrix} \quad , \quad \pm \begin{bmatrix} -1 \\ 1 \\ 1 \\ 1 \end{bmatrix}, \quad \pm \begin{bmatrix} 0 \\ 1 \\ 0 \\ 1 \end{bmatrix}, \quad \pm \begin{bmatrix} 1 \\ 0 \\ -1 \\ 1 \end{bmatrix}$$

$$\underbrace{\qquad\qquad\qquad\qquad}_{q(\cdot)=\pm 1} \qquad \underbrace{\qquad\qquad\qquad\qquad}_{q(\cdot)=0} \qquad (26)$$

Thus, by virtue of Eq. (22), the first 6 vectors belong to the orbit $P_1$ and the other 6 vectors do not belong to any pitch orbit, so only $P_1$ intersects nontrivially the set $E$. It can be checked, by direct calculation, that postulate O2 is also satisfied.

Let us return to postulate M i.e. definition of characteristics $F^\#$. Using notation of this section we may rewrite Eq. (13) in the form:

$$F^\#(S) = \sum_{w \in S} \log^+ \left\| F_u \otimes F_v \otimes F_w \otimes F_t(w) \right\| \qquad (27)$$

It still remain the problem to determine $F_u, F_v, F_w$ and $F_t$. Let us recall the physiological argument that each row of the hair cells provides spectral information to the particular coordinate in the spectral basis of the space $W$. It is shown schematically in Fig.4. Investigations of the ear physiology show that the outer hair cells are more sensitive in comparison with the inner hair cells, which are more passive detectors (Boer, 1991, Sect.7.7). Thus in our model, at low level signals we make a simplification[7] that all information originates effectively from three rows of outer hair cells, which corresponds to three dimensions determined by the coordinates $u$, $v$, $w$. If we take into account this assumption, we may introduce the following characteristics:

---

[7] This simplification will be discussed also in Sect. VI.

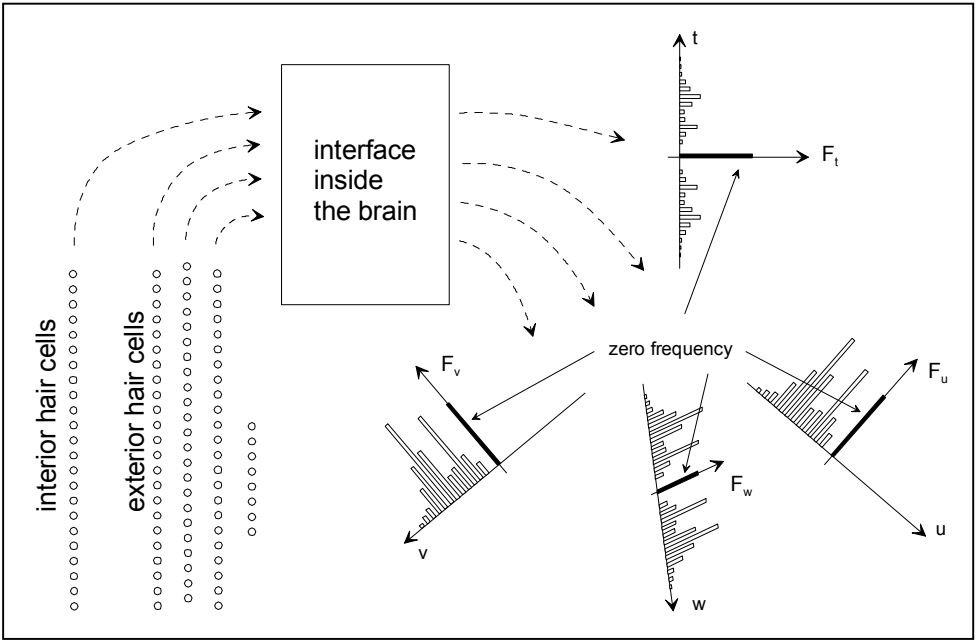

Fig. 4. Scheme of joining of the information provided from inner ear with the considered model. Three rows of outer hair cells and one row inner hair cells provide spectral characteristics.

$$F_u := \begin{cases} 0 & \text{for} \quad \omega > 0 \\ 10 & \text{for} \quad \omega = 0 \\ F(\omega) & \text{for} \quad \omega < 0 \end{cases} \qquad F_v := \begin{cases} F(\omega) & \text{for} \quad \omega > 0 \\ 10 & \text{for} \quad \omega = 0 \\ 0 & \text{for} \quad \omega < 0 \end{cases}$$

$$F_w := \begin{cases} F(\omega) & \text{for} \quad \omega > 0 \\ 10 & \text{for} \quad \omega = 0 \\ F(\omega) & \text{for} \quad \omega < 0 \end{cases} \qquad F_t := \begin{cases} 0 & \text{for} \quad \omega > 0 \\ \frac{1}{\max(F)} & \text{for} \quad \omega = 0 \\ 0 & \text{for} \quad \omega < 0 \end{cases}$$

$$(28)$$

where $max(F)$ is the maximal absolute value of the set, which consists of the all values of $F$ and the number 1. Let us comment now some important elements in the above definitions. In an artificial way the zero frequency component was introduced with value 10 i.e. 10dB in definitions $F_u$, $F_v$, $F_w$. This operation amplifies sensation of weak signals. By virtue of the mentioned simplification, we neglect all nonzero frequency components of $F_t$. The term $1/\max(F)$ in definition $F_t$ decreases the characteristics $F^{\#}$ by multiple of number $\log(\max(F))$. As a result we obtain the rise of the threshold of hearing, which may be interpreted as a kind of masking mechanism. $F_u$ depends only on negative frequencies and $F_v$ depends only on positive frequencies. This causes the characteristics $F^{\#}$ to depend on the order structure of the field $K$. In consequence the characteristics $F^{\#}$ treats differently high and low frequencies what is in agreement with observation. In the next section we describe combination tones phenomenon in the range of low-level signals with the help of the above concepts.

## V. COMBINATION TONES

As we mentioned in the introduction, combination tones are observed with the signals consisting of two frequency components. If the frequencies of these components are $\omega_1, \omega_2$ and $\omega_2 > \omega_1$, the pitch of the combination tones cor-responds to frequencies $a\omega_1 + b\omega_2$ where $a$, $b$ are small integer numbers. The most prominent among them is the combination tone $2\omega_1 - \omega_2$, which is audible at a low level of primary tones. In order to describe this phenomenon we consider a signal consisting of two sine components $\omega_1, \omega_2$ ($\omega_2 > \omega_1 > 0$) subjected to NMT of only second order. We will assume for simplification that all the components of the spectrum have the same level equal to 10 dB and the same initial phase equal to $\pi/2$ (the latter simplifies a notation only). This slightly idealised case will help one to understand the example model. For clarity we write down $F$ in the form:

$$F = F_0 + \Delta \qquad (29)$$

where

$$F_0 := 10(\delta_{-\omega_1} + \delta_{\omega_1} + \delta_{-\omega_2} + \delta_{\omega_2}) \qquad (30)$$

$$\begin{aligned} \Delta := 10(&\delta_{-2\omega_1} + \delta_{2\omega_1} + \delta_{-2\omega_2} + \delta_{2\omega_2} + \\ &\delta_{-\omega_2-\omega_1} + \delta_{\omega_2+\omega_1} + \delta_{\omega_1-\omega_2} + \delta_{\omega_2-\omega_1}) \end{aligned} \qquad (31)$$

and $\delta_{\omega} : K \to C$ such that

$$\delta_{\omega}(\xi) := \begin{cases} 1 & \text{for} \quad \xi = \omega \\ 0 & \text{for} \quad \xi \neq \omega \end{cases} \qquad (32)$$

$F_0$ is the spectrum of the primary tones and $\Delta$ is the part of the spectrum describing the nonlinear distortion. Our goal is to find the distribution of the pitch $F^{\#}(P_\omega)$ depending on $\omega_1$ and $\omega_2$. We will achieve this in three steps. In the first step we will find those elements of the space $W$ on which the function $F_u \otimes F_v \otimes F_w \otimes F_t$ from Eq. (27) has nonzero values. Since a subset on which $F$ has nonzero values is the set

$$\Omega := \{-\omega_1, -\omega_2, -\omega_1-\omega_2, \omega_2-\omega_1, -2\omega_1, -2\omega_2, \omega_1, \omega_2,$$
$$\omega_1+\omega_2, \omega_1-\omega_2, 2\omega_1, 2\omega_2\} \qquad (33)$$

then according to the definitions (28) of the characteristics $F_u, F_v, F_w, F_t$, the set on which $F_u \otimes F_v \otimes F_w \otimes F_t$ has non-zero value is

$$Z := \{-\omega_1, -\omega_2, -\omega_1-\omega_2, \omega_2-\omega_1, -2\omega_1, -2\omega_2, 0\} \times$$
$$\times \{0, \omega_1, \omega_2, \omega_1+\omega_2, \omega_1-\omega_2, 2\omega_1, 2\omega_2\} \times \qquad (34)$$
$$\times \{-\omega_1, -\omega_2, -\omega_1-\omega_2, \omega_2-\omega_1, -2\omega_1, -2\omega_2, 0,$$
$$\omega_1, \omega_2, \omega_1+\omega_2, \omega_1-\omega_2, 2\omega_1, 2\omega_2\} \times \{0\}$$

We pass to the second step. Since the pitch orbits are included in $W_0$ (see IV.C) we have to check whether the given $z \in Z$ belongs to subspace $W_0$. Each element $z \in Z$ can be written as a linear combination of variables $\omega_1, \omega_2$:

$$z = \omega_1 z_1 + \omega_2 z_2 \qquad (35)$$

where $z_1, z_2 \in W$. The membership of $z$ to $W_0$ is verified by the use of the condition (24), which can be written in the form

$$\varphi(z) = 0 \qquad (36)$$

where $\varphi$ is a linear form, which is equal to $[2, 1, 1, -1]$ in the dual basis to the spectral basis, and using Eq. (35) we have

$$\omega_1 \varphi(z_1) + \omega_2 \varphi(z_2) = 0 \qquad (37)$$

If $\varphi(z_1) = \varphi(z_2) = 0$ then above equation is satisfied for every $\omega_1, \omega_2$, and such an element $z \in Z$ will be called *unconditional*. If $\varphi(z_1) \neq 0$ and $\varphi(z_2) \neq 0$ then Eq. (37) has solutions for a certain ratio $\omega_2 : \omega_1$, and if $\omega_2 : \omega_1 > 1$ then such an element $z \in Z$ will be called *conditional*. In the remaining cases proper solutions of Eq. (37) are not obtained. The subset of the given pitch orbit $P_\omega$ containing all unconditional elements of the set $Z$ will be called *unconditional component of the orbit $P_\omega$*. Similarly, the subset of the given pitch orbit $P_\omega$ containing all conditional elements of the set $Z$ for some ratio $\omega_2 : \omega_1$ will be called *conditional component of the orbit $P_\omega$ for ratio $\omega_2 : \omega_1$. We will see in a moment that unconditional components represent real tones, which occur in the sound spectrum, while conditional components are perfect candidates for combination tones.*

In the third step we determine the pitch orbit which any element $z \in Z \cap W_0$ belongs to, at a given $\omega_1$ and $\omega_2$. Since $z \in W_0$, it is sufficient to use form $q$. At $t = 0$ using Eq. (25) we have

$$q([u,v,w,0]) = -(u+v)^2 \qquad (38)$$

TABLE I. List of all unconditional elements and some conditional elements for $\omega_2 : \omega_1 = 4 : 3$ of the set $Z$.

| No. | Vector coordinates (t = 0) | | | Frequency of the pitch orbit $\|u+v\|$ | $\omega_1, \omega_2$ for which: $2u+v+w=0$ |
|---|---|---|---|---|---|
| | $u$ | $v$ | $w$ | | |
| 1 | $0$ | $\omega_1$ | $-\omega_1$ | $\omega_1$ | For all $\omega_1, \omega_2$ |
| 2 | $-\omega_1$ | $0$ | $2\omega_1$ | | |
| 3 | $-\omega_2$ | $\omega_1+\omega_2$ | $\omega_2-\omega_1$ | | |
| 4 | $-\omega_2$ | $\omega_2-\omega_1$ | $\omega_1+\omega_2$ | | |
| 5 | $-\omega_1$ | $2\omega_1$ | $0$ | | |
| 6 | $0$ | $\omega_2$ | $-\omega_2$ | $\omega_2$ | |
| 7 | $-\omega_2$ | $2\omega_2$ | $0$ | | |
| 8 | $-\omega_2$ | $0$ | $2\omega_2$ | | |
| 9 | $-\omega_1$ | $\omega_1+\omega_2$ | $\omega_1-\omega_2$ | | |
| 10 | $0$ | $\omega_2-\omega_1$ | $\omega_1-\omega_2$ | $\omega_2-\omega_1$ | |
| 11 | $-\omega_1-\omega_2$ | $2\omega_2$ | $2\omega_1$ | | |
| 12 | $-\omega_1-\omega_2$ | $2\omega_1$ | $2\omega_2$ | | |
| 13 | $0$ | $\omega_1+\omega_2$ | $-\omega_1-\omega_2$ | $\omega_1+\omega_2$ | |
| 14 | $\omega_1-\omega_2$ | $2\omega_2$ | $-2\omega_1$ | | |
| 15 | $0$ | $2\omega_1$ | $-2\omega_1$ | $2\omega_1$ | |
| 16 | $0$ | $2\omega_2$ | $-2\omega_2$ | $2\omega_2$ | |
| 17 | $-2\omega_1$ | $\omega_2$ | $2\omega_2$ | $2\omega_1-\omega_2$ | If $\omega_2 : \omega_1 = 4 : 3$ |
| 18 | $\omega_1-\omega_2$ | $\omega_1$ | $\omega_1-\omega_2$ | | |
| 19 | $-2\omega_1$ | $2\omega_2$ | $\omega_2$ | $2\omega_2-2\omega_1$ | |
| 20 | $\omega_1-\omega_2$ | $2\omega_1$ | $-\omega_2$ | $3\omega_1-\omega_2$ | |

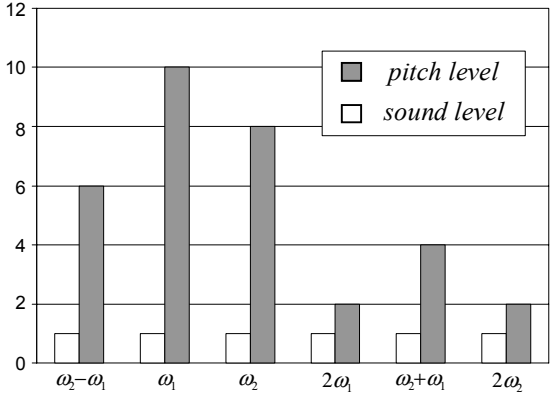

Fig. 5. The pitch distribution for the spectrum $F$, which is constant on the set $\Omega$ given by Eq. (33) and has value 10dB. There are only the unconditional pitch components. Such components correspond to real tones i.e. components of the spectrum $F$. $F^{\#}$ favours pitches which correspond to primary tones. The sound level is given by $\log^{+}(|F(\omega)|)$ and the pitch level is given by $F^{\#}(P_{\omega})$.

Moreover Eq. (22) leads to the conclusion that if $[u,v,w,0] \in W_0$ then $[u,v,w,0] \in P_\omega$ under condition that $u+v = \pm\omega$. All unconditional elements of the set $Z$ (except 0) and conditional elements for ratio $\omega_2 : \omega_1 = 4:3$ are shown in Table I. Figure 5 shows pitch distribution for the considered spectrum $F$ restricted to unconditional components only. We observe, that the presented mechanism amplifies the pitches that correspond to prime tones. Let us now proceed to conditional components shown in Table I. Let us note that an additional pitch appears for $\omega_2 : \omega_1 = 4:3$, which represents tone with frequency $2\omega_1 - \omega_2$, although there is no such a component in the spectrum. Moreover, for this frequency ratio we have the following equalities:

$$\begin{aligned} |2\omega_1 - \omega_2| &= |2\omega_2 - 2\omega_1| \\ |3\omega_1 - \omega_2| &= |2\omega_2 - \omega_1| \end{aligned} \qquad (39)$$

Therefore, for ratio $\omega_2 : \omega_1 = 4:3$ we have only two conditional pitch components: the first one contains vectors #17,18,19 and represents frequency $2\omega_1 - \omega_2$, the second contains the vector #20 and represents the frequency $2\omega_2 - \omega_1$. Let us note, that the second conditional pitch component is considerably smaller than the first one. Moreover, we will see in a moment that this pitch component in most cases is eliminated due to masking properties of $F^{\#}$. It can be directly checked that there is no such frequency ratio, for which the set $Z$ intersects the pitch orbits, which represent sum frequencies $2\omega_1 + \omega_2$ and $2\omega_2 + \omega_1$. We conclude, that our method allows one to explain these experimental facts, which are quite difficult to understand on the base of the Helmholtz theory (Boer 1984, Plomp 1965).

Figure 6 presents frequency ratios $\omega_2 : \omega_1$ of combination tones in the range from 1 to 2, which are predicted by the model[8]. Apart from nonlinear distortion of the 2nd order, the nonlinear distortions of the 3rd and the 4th order in the spectrum $F$ were also taken into account. If we compare the diagram in Figure 6 with that by Plomp (1965) we observe compatibility with regards to type of occurrence of combination tones. The types $2\omega_1 - \omega_2$, $3\omega_1 - 2\omega_2$ and $4\omega_1 - 3\omega_2$ are clearly distinguished. What is very interesting is that, they already originate from NTM's of the 2nd, the 3rd and the 4th order respectively. However, in comparison with experiment, this model predicts too few combination tones for $\omega_2 : \omega_1 < \frac{3}{2}$ and too many for $\omega_2 : \omega_1 > \frac{3}{2}$. It also predicts to many combination tones over $\omega_1$ frequency (Plomp, 1965).

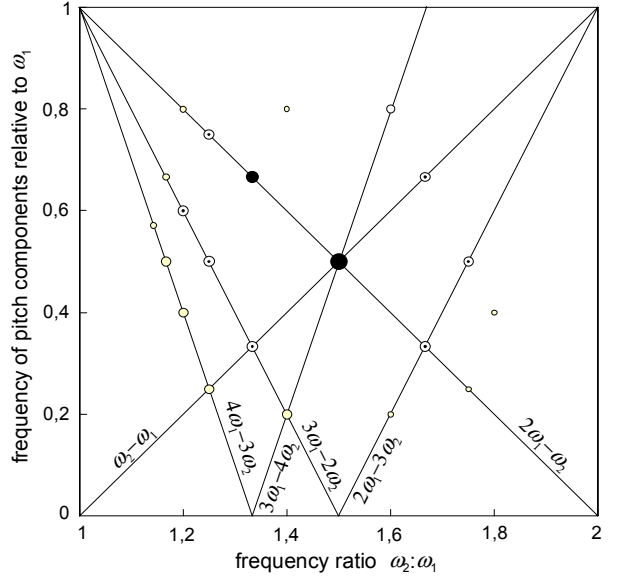

Fig. 6. Relative frequencies ($\omega_2 : \omega_1$) of combination tones (conditional pitch components) predicted by the model in the range of 1 to 2. Area of circle represents relative magnitude of the combination tone. Symbol "●" denotes that combination tone, which originates from NMT's of the 2nd, the 3rd, and the 4th order and similarly "◉" from NMT's of the 3rd and the 4th order and "○" from NMT's of the 4th order only.

The characteristics $F^{\#}$ in Eq. (27) has masking features. We will illustrate this on another example of spectrum $F$ in the form $F = F_0 + \Delta$, where

$$F_0 := 10^3 (\delta_{\omega_1} + \delta_{-\omega_1}) + 10(\delta_{\omega_2} + \delta_{-\omega_2}) \qquad (40)$$

and $\Delta$ is the same as before, Eq. (31). The Fig.7 shows results of using characteristics $F^{\#}$ to this spectrum. We observe that the unconditional components representing nonlinear distortion were masked. We note that $F^{\#}$ masks the nonlinear components and does not mask $\omega_2$ component, in spite of the fact that spectrum $F$ contains all of them at the same level. Moreover, if $\omega_2 : \omega_1 = 4:3$ then in contrary to the $2\omega_2 - \omega_1$ component, the $2\omega_1 - \omega_2$ component is not masked, what is experimentally observed. This masking mechanism works when nonlinear distortions are at least 10-20 dB smaller then the maximal prime component. However even at comparable levels, there is an effect of suppression of at last all nonlinear components with higher frequencies. It should be emphasized, that proposed masking mechanism does not completely explain the known masking effect. We should remember that the auditory process is a multi-level phenomenon, and one needs to take into account the known masking mechanisms at the level of the human ear (Moore 1997).

[8] The model predicts also combination tones for $2 \le \omega_2 : \omega_1 \le 11$. There are also predicted combination tones with frequencies above $\omega_1$ frequency. Nevertheless, there is the highest density and the highest relative magnitude of combination tones in the region presented on Fig. 6.

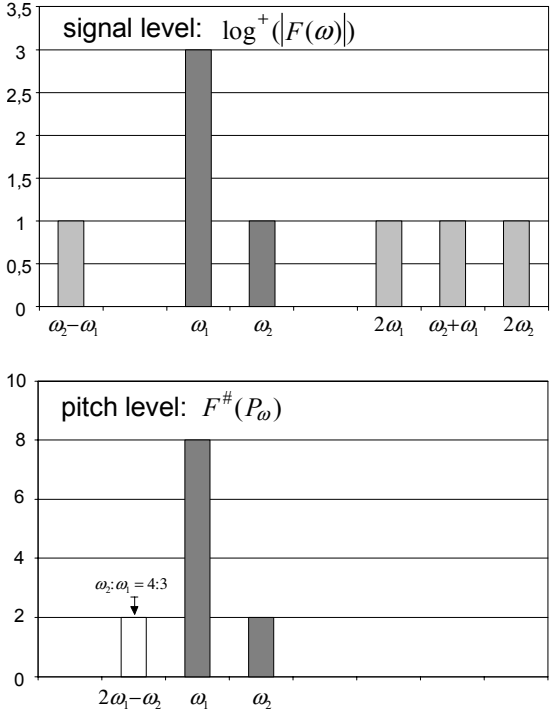

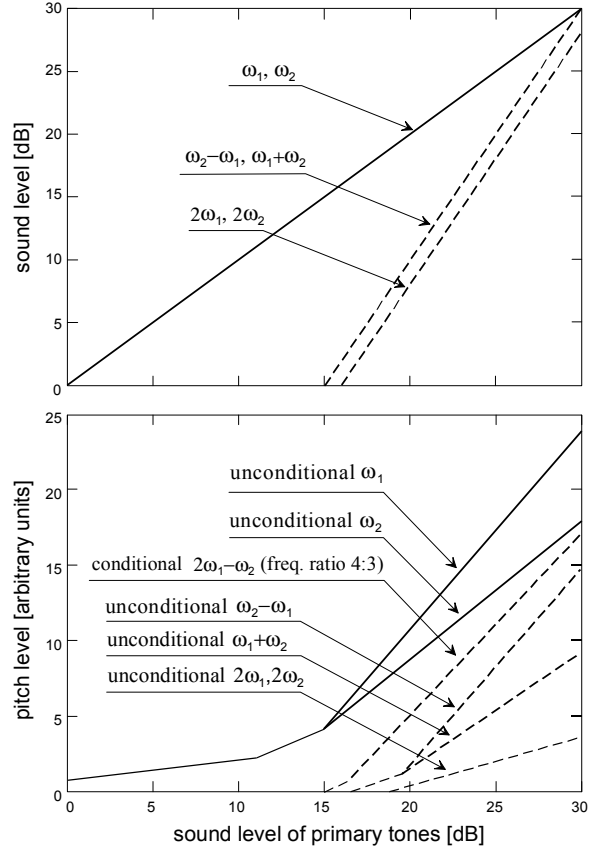

Fig. 7. The masking effect. $F$ is given by Eqs. (31) and (40). $F^{\#}$ masks the unconditional pitch components corresponding to the nonlinear tones. The combination tone $2\omega_1 - \omega_2$ (conditional pith component) for $\omega_2 : \omega_1 = 4 : 3$ is not masked.

Fig. 8. Sound level (upper diagram) and pitch level (lower diagram) of the signal consisting of two sine components given by Eq. (43) subjected to the NMT of the second order given by Eq. (42). The ordinate gives the sound level of the primary tones, which is assumed to be the same for both of tones.

Now we will study dependence of the level of the pitch components on the sound level. In these considerations we follow Plomp (1965) and Boer (1984). To avoid complications of nonlinearity combined with frequency selectivity let us consider device that transforms signal $x(t)$ into signal $y(t)$ according to equation

$$y(t) = T(x(t)) \tag{41}$$

were $T$ is the NMT of the second order given by

$$T(u) := u + \varepsilon u^2 \tag{42}$$

We consider a two-component signal in which both components have the same amplitude

$$x(t) := A(\cos \omega_1 t + \cos \omega_2 t) \tag{43}$$

The spectral characteristics $F_x$ and $F_y$ of signals $x(t)$ and $y(t)$ respectively are

$$F_x = A F_0 \qquad \text{and} \qquad F_y = A F_0 + \varepsilon A^2 \Delta \tag{44}$$

where this time

$$F_0 := \delta_{-\omega_1} + \delta_{\omega_1} + \delta_{-\omega_2} + \delta_{\omega_2}$$

$$\Delta := \delta_{-\omega_2 - \omega_1} + \delta_{\omega_2 + \omega_1} + \delta_{\omega_1 - \omega_2} + \delta_{\omega_2 - \omega_1} +$$
$$\tfrac{1}{2}(\delta_{-2\omega_1} + \delta_{2\omega_1} + \delta_{-2\omega_2} + \delta_{2\omega_2}) + 2\delta_0 \tag{45}$$

Figure 8 shows result of calculation of $\log^+(F_y)$ -upper part and $F_y^{\#}$ - lower part depending on $\log(A)$, with $\varepsilon = 10^{-3}$. We observe that pitch level of the nonlinear tones grows slower than the pitch level of primary tones. There is only one exception namely the pitch level of $\omega_2 - \omega_1$ tone grows slightly faster than $\omega_2$ tone. In contrast, the level of the sound of all nonlinear tones grows twice as fast as the level of the sound of the primary tones. The most interesting is the fact that the pitch level of combination tone $2\omega_1 - \omega_2$ grows as quickly as pitch level of the primary tone $\omega_1$. This is in agreement with experiment (see introduction).

## VI. FINAL REMARKS

The example model presented in the previous section is very much simplified and does not solve many other problems. It should be rather treated as a starting point for the development of more effective models. We draw our attention to only two assumptions, which have been made in this model.

The dimension of the space $W$ plays an important role in simulating of NMT's. If we have more coordinates we will have more ability to generate derivative vector (see Sect. II). For example if we have 3D vector with frequencies $\omega_1, \omega_2, \omega_3$ as coordinates then we can obtain derivative vector with $3^{rd}$ order nonlinear tone of frequency $\omega_1 + \omega_2 + \omega_3$. Such a case is impossible in a 2D model. The example model presented in Section IV looks as if a 4D model but it is in fact a 3D one because measure $F^\#$ given by Eq. (27) is "3D". A more effective model should be fully 4D. In consequence we will obtain also higher number of combination tones. Therefore, the assumption in subsection IV.E that all nonzero frequency components of $F_t$ may be neglected at low level signals, is rather too strong simplification.

Experiments show that the phase relations between components of the signal play a certain role in the psychoacoustic phenomena. For simplicity, we did not take into account the phase in the presented examples. However, it could be included by assuming a different "norm" e.g. $\| \cdot \| := |\mathrm{Re}(\cdot)|$ in Eq. (2). By proper adjustment of the "norm" in Eq. (2) the dependence of sensation level on sound level for weak signals can be also improved. (see Fig. 8). For instance, a "norm" with upper bound may be considered. The correctness of any solution should be verified experimentally.

## VII.  CONCLUSIONS

The presented theory of sensation analysis allows one to understand the occurrence of combination tones and to explain experimental facts that are inexplicable within the Helmholtz theory. In particular, it was shown, that the combination tones of the type $\omega_1 - n(\omega_2 - \omega_1)$ can originate from nonlinear distortions of the order $n+1$, where $n = 1,2,3$. This means that the tone $2\omega_1 - \omega_2$ can originate from nonlinear distortions of the $2^{nd}$ order and therefore it is audible at relatively low levels of the sound. Within the range of low level signals presented model explains the following experimental facts.

- The number of combination tones is considerably lower than potential nonlinear tones. Only few of them occur among possible types of combination tones and they appear at some determined frequency ratios of primary tones.
- The tones $2\omega_1 + \omega_2$ and $2\omega_2 + \omega_1$ do not accompany the audible tone $2\omega_1 - \omega_2$. Apart from this, the tone $2\omega_2 - \omega_1$ is not audible at low level signals.
- The level of the combination tone $2\omega_1 - \omega_2$ grows approximately as quickly as the level of the pitch of the primary tones.

However, in comparison with experiment, presented model predicts too few combination tones for $\omega_2 : \omega_1 < \frac{3}{2}$ and too many for $\omega_2 : \omega_1 > \frac{3}{2}$. It also predicts to many combination tones over $\omega_1$ frequency. These difficulties may be overcome in fully 4D model.

## AKNOWLEDGMENTS

The author wishes to express special thanks to Krzysztof Pachucki for the content-related help to prepare this paper and also to Adam Podhorski and Gary Stewart for language corrections of the manuscript.